\def\year{2022}\relax
\documentclass[letterpaper]{article}

\usepackage{amsmath}
\usepackage{amssymb}
\usepackage{bbm}
\usepackage{xcolor}
\usepackage{hyperref}
\usepackage[switch]{lineno}  
\usepackage[caption=false]{subfig}

\usepackage{aaai22} 
\usepackage{times} 
\usepackage{helvet} 
\usepackage{courier} 
\usepackage{graphicx} 
\def\UrlFont{\rm} 
\usepackage{natbib} 
\usepackage{caption} 
\DeclareCaptionStyle{ruled}%
{labelfont=normalfont,labelsep=colon,strut=off}
\title{Multi-Agent Dynamic Pricing in a Blockchain Protocol Using Gaussian Bandits}
\author{Alexis Asseman, Tomasz Kornuta, Anirudh Patel, Matt Deible, Sam Green}
\affiliations{
\{alexis,tomasz,anirudh,matt,sam\}@semiotic.ai\\
Semiotic Labs\\
Los Altos, CA, US}
\copyrighttext{}

\begin{document}
\maketitle
\pagenumbering{arabic}
\begin{abstract}
The Graph Protocol indexes historical blockchain transaction data and makes it available for querying.
As the protocol is decentralized, there are many independent Indexers that index and compete with each other for serving queries to the Consumers.
One dimension along which Indexers compete is pricing.
In this paper, we propose a bandit-based algorithm for maximization of Indexers' revenue via Consumer budget discovery.
We present the design and the considerations we had to make for a dynamic pricing algorithm being used by multiple agents simultaneously.
We discuss the results achieved by our dynamic pricing bandits both in simulation and deployed into production on one of the Indexers operating on Ethereum.
We have open-sourced both the simulation framework\footnote{\url{https://github.com/semiotic-ai/autoagora-agents}} and tools\footnote{\url{https://github.com/semiotic-ai/autoagora}} we created, which other Indexers have since started to adapt into their own workflows.
\end{abstract}

\section{Introduction}

Blockchain~\cite{zheng2018blockchain} technology is one of the most promising technologies for the next generation of applications, from decentralized finance and money transfers~\cite{tapscott2017blockchain}, to secure multi-party computation via smart contracts~\cite{wood2014ethereum,buterin2014next}, to secure supply chain management~\cite{saberi2019blockchain}.


At the core of blockchain technology, there is a public, distributed ledger -- a chain of immutable ``blocks''.
The chain continuously grows when new blocks are created and appended to it.
Smart contract-based blockchains like Ethereum change state through transactions. As the information related to specific applications is typically scattered throughout different blocks and transactions, searching for information in distributed ledgers is not an easy task~\cite{third2017linked}.

The Graph protocol~\cite{thegraph2022} is addressing this problem by providing open-source software for indexing and querying distributed ledgers in a decentralized manner.
The Graph's developers build and publish open APIs, called subgraphs, that Consumers can later query using GraphQL~\cite{graphql2015spec}.

\begin{figure*}[htbp]
    \centering
    \includegraphics[width=0.7\textwidth]{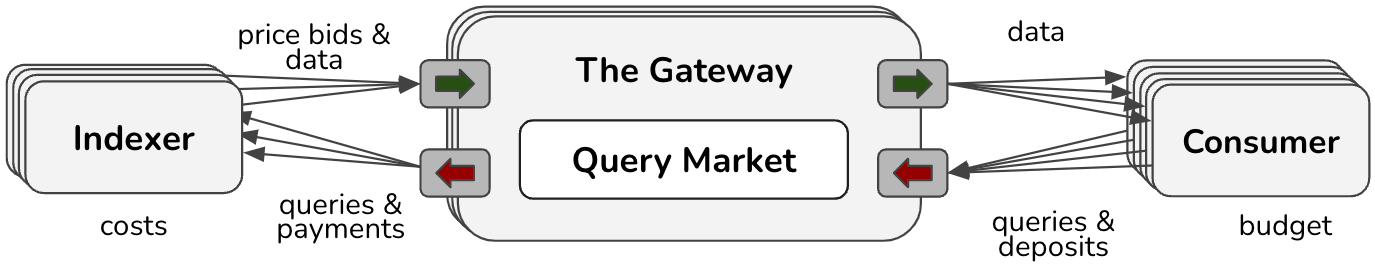}
    \caption{A simplified scenario for query serving in The Graph protocol}
    \label{fig:scenario}
\end{figure*}

\autoref{fig:scenario} presents a simplified Graph scenario, with many independent entities called \textbf{Indexers} (data sellers), who both operate indexing nodes and compete to sell queries to \textbf{Consumers} (data buyers).
It also includes \textbf{Gateways} that constitute entry points for Consumers who submit a query and budget to one of the Gateways. The Gateway then collects price bids from Indexers for the submitted query.
Based on various metrics, including Indexer price, data freshness, and historical latency, the Gateway selects an Indexer to serve the query and payment and, once the Indexer returns the results, passes them back to Consumer.

Note the primary goal of The Graph protocol is to provide a high Quality of Service (QoS), which, in the decentralized context, means that it is in the best interest of Consumers to attract more than one Indexer to a given subgraph.
On the other hand, the Consumer budget for a given query is not publicly revealed.
As a result, Indexers, Consumers, and Gateways form a game with imperfect information, wherein different entities are driven by different incentives.

\subsection{Our Contributions}
In this work, we have focused on the problem of maximization of an Indexer's profits via dynamic pricing~\cite{nakhe2017dynamic}.
As estimation of infrastructure costs (i.e. executing GraphQL queries~\cite{hartig2018semantics,mavroudeas2021learning}) is difficult, we decided to use Reinforcement Learning~\cite{sutton2018reinforcement} for the dual problem of maximizing revenue by consumer budget discovery.
The contributions of the paper are the following:
\begin{itemize}
    \item Application of sample-efficient Gaussian bandits with continuous price action space for dynamic pricing,
    \item Simulation-based verification that our solution works properly in a zero-sum game.
\end{itemize}
The latter is crucial, as the solution is to be deployed in production by multiple Indexers.
Moreover, we discuss how our bandits reveal the importance of well-designed Gateways, accompanied by experiments showing catastrophic outcomes with improper design.
Finally, we show the results of deploying our solution in production in one of the Indexers serving data on the second largest blockchain, Ethereum.

\section{Related Work}

As Multi-Agent Reinforcement Learning (MARL)~\cite{zhang2021multi} draws heavily from game theory, dynamic pricing problems have been an active area of applied research for MARL.
Tesauro et al. were the first to apply reinforcement learning in this domain \cite{tesauro2002pricing}.
They applied simultaneous Q-learning to two competing agents.
Each agent competes only on the basis of price.
While they showed good results in the case in which one agent learns and the other remains fixed, the case in which both agents were allowed to learn was less likely to converge.
The obvious drawback to Tesauro et al.'s approach was their choice to model the problem using an approach similar to independent Q-learning (IQL) \cite{tan1993multi}.
In practice, IQL has been shown to work quite well for certain classes of problems \cite{matignon2012independent}.
However, (simultaneous) IQL also makes the environment non-stationary from any individual agent's perspective.
Thus, it negates convergence guarantees \cite{foerster2017stabilising}.

Naturally, we could extend Tesauro et al. by applying concepts from multi-agent reinforcement learning.
In particular, note that the simultaneity of their approach worsens the convergence guarantees significantly.
K\"on\"onen makes this observation and instead tries to solve the same problem using asymmetric multi-agent reinforcement learning \cite{kononen2006dynamic}.
They modeled the dynamic pricing problem as a Markov Game.
They then applied both a policy gradient method and a Q-learning method to the problem, ultimately finding that both methods achieved similar profits.
Whilst this approach did achieve good results, it does require that each agent ``announces'' its price decision to the other agent.
In cases of limited observability, such as in The Graph protocol, such an assumption does not hold.

In the case of The Graph protocol, we also want to limit online price experimentation, as each updated price from our agent is slow to propagate through the system.
Here, Cheung et al. discovered that they could use regret as a way to characterize the effect of bounding the number of price changes \cite{cheung2017dynamic}.
For a sales horizon of $T$ periods with at most $m$ price changes, their policy has a regret bounded by $O(\log^{(m)}T)$.
This approach was actually so successful that its application in Groupon increased their daily bookings by 116\% and their daily revenue by 21.7\%.

Maestre et al. have also applied reinforcement learning to the problem of dynamic pricing \cite{maestre2018reinforcement}.
They noted that while traditional dynamic pricing is focused on revenue maximization, in a practical system, there ought to be some notion of ``fairness'' that balances against revenue.
They use Jain's index \cite{jain1984quantitative} as a metric for fairness.

In addition to Groupon, Alibaba has also implemented reinforcement learning in production for dynamic pricing \cite{liu2019dynamic}.
Liu et al.'s primary contribution was the introduction of a new type of reward function -- the difference of revenue conversion -- rather than the standard revenue.
They also use pre-training on historical data to give their deep reinforcement learning algorithm a warm start.
Ultimately, we do not have access to such historical data and are therefore unable to replicate their pre-training method.

Broadly speaking, major contributions in applying reinforcement learning to dynamic pricing have primarily come from single-agent environments.
Some of the earlier literature takes a more game theoretical approach to try to understand what happens when two agents run the same optimization algorithm against each other.
Our application lies in the middle of these two.
We need to consider the multi-agent case as our tool will be adopted by multiple Indexers participating in the same marketplace.
At the same time, we want to apply many of the lessons of modern research efforts, particularly those that try to solve similar problems like Cheung et al.
This intersection means that we need to take a somewhat novel approach.

\begin{figure*}[htbp]
    \centering
    \subfloat[]{
        \includegraphics[width=0.65\textwidth]{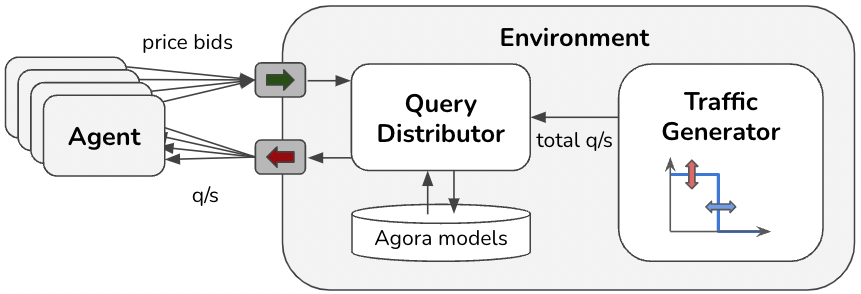}
        \label{fig:simulation_overview}
    }
    \hfill
    \subfloat[]{
        \includegraphics[width=0.26\textwidth]{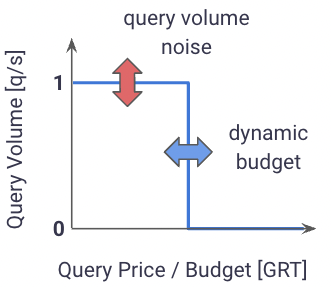}    
        \label{fig:simulation_traffic_gen}
    }
    \caption{(a) The Graph scenario modeled in our experiments, with the \textbf{Environment} consisting of components generating the traffic and distributing the queries depending on the price-bids. (b) An illustration of query volume as a function of query price, with the step representing the Consumer's budget per query. If an Indexer's price bid is greater than the budget, the Indexer will not be selected to serve the query.}
    \label{fig:simulation}
\end{figure*}

\section{Problem Overview}
As The Graph is a decentralized protocol, we do not provide a way for any given Consumer to select any given Indexer.
Instead, Consumers specify preferences (such as latency, reliability, budget, etc.), and one of Gateway component, called the \textbf{Indexer Selection Algorithm (ISA)}, matches an Indexer to a Consumer.
In a sense, the ISA is fundamentally just a matching algorithm.
Though the ISA considers multiple factors, the only knob, so to speak, that an Indexer can tweak, is its price.
If an Indexer says it will serve a query for a price that is within a Consumer's budget, the Indexer \textit{may} be selected to serve that query.

Practically speaking, this is still an oversimplification.
Different queries have different infrastructure costs to Indexers, so an Indexer may choose to set its price to be higher for certain queries and lower for others.
Moreover, different indexers may have different infrastructure costs for serving the same query.
Add to this the fact that there are a potentially infinite number of different query types.
As this quantity is fundamentally unknowable to us, and it is nearly impossible to model,
we decided to not take this into account.

\section{Offline Verification}

It would be irresponsible of us to deploy a solution to production without first verifying its behavior offline.
The following section details the results of those experiments.

\subsection{Experimental Setup}
The environment consists of two major components: the \textbf{Query Distributor} and the \textbf{Traffic Generator}, shown in \autoref{fig:simulation_overview}.
The \textbf{Query Distributor}, being the wrapped, aforementioned \textbf{ISA}, collects the price-bids of competing Indexer agents (expressed in The Graph's domain-specific language -- Agora) and splits the total query volume between the agents by looking at their price-bids, QoS, etc.
The \textbf{Traffic Generator} simulates the queries sent by the Consumers.
In each step, it generates a normalized query volume with additional noise (additive white Gaussian noise represented by the red arrow in \autoref{fig:simulation_traffic_gen}).
Moreover, the environment models the fact that Consumers have a limited budget that can change over time (the blue arrow in \autoref{fig:simulation_traffic_gen}).
If the price bid of a given agent goes beyond the budget, the agent won't receive any queries at a given time step.

\subsection{Single-Agent Experiments}

At a high level, in this section we will consider the single-agent decision problem and validate whether an isolated agent (i.e. without competition from other agents) can select prices so as to maximize revenue.
In this case revenue maximization means setting up prices as close to the (unobservable) Consumer budget\footnote{In The Graph each Consumer sets a separate budget for each query. Within the paper we will refer to it simply as budget.} as possible, hence we validate whether an agent can successfully discover the budget.

The agent chooses some price $p_t\in\mathcal{P}$.
Here, $\mathcal{P}$ is the set given by the continuous interval $[\underline{p}, \overline{p}]$.
The vector of all budgets at the current time step is $\textbf{b}_t$.
For a given query with budget $b^j_t$ (the $j$th element of $\textbf{b}_t$) the ISA will select the agent with probability 1 if $p_t \leq b^j_t$.\footnote{Recall, this is the single-agent decision problem, so the ISA has no choice but to select this agent.}
If the agent selects $p_t > b_t^j$, the ISA will not select the agent.
Thus, the agent's reward would be

\begin{equation}
    \label{eq:sarl-reward}
    r_t(p_t, \textbf{b}_t) = \sum_{j} p_t\mathbbm{1}\{p_t \leq b^j_t\}
\end{equation}

\begin{figure}[t!]
    \centering
    \subfloat[Step 64: shifting from initial action distribution]{
        \includegraphics[width=0.45\textwidth]{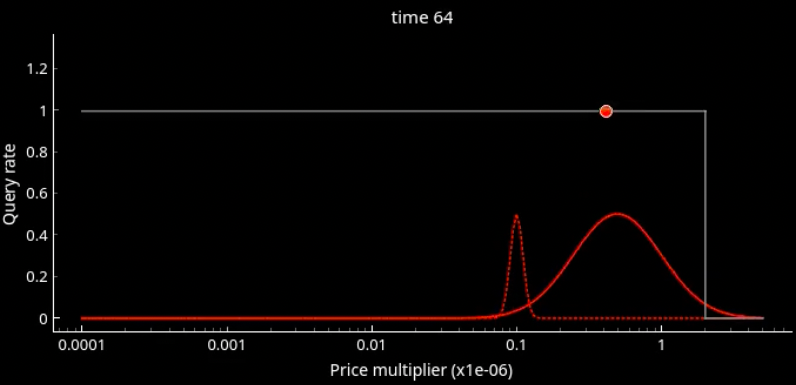}
        \label{fig:1ppo_fixed_budget_init}
    }\\
    \subfloat[Step 999: policy reached suboptimal distribution]{
        \includegraphics[width=0.45\textwidth]{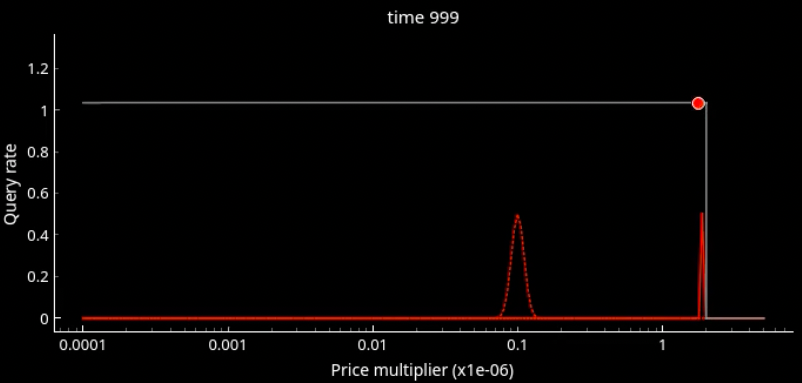}
        \label{fig:1ppo_fixed_budget_subopt}
    }
    \caption{Experiments with fixed Consumer budget and no competition, showing how bandit's policy evolves (a) and converges to suboptimal distribution (b).
    Red dashed lines represent initial policy distributions, red solid lines indicate current policy distribution. Red dots indicate query volumes served by the bandit mapped to its price bids. Total query volumes and Consumer budgets are indicated by white lines.} 
    \label{fig:1ppo_fixed_budget}
\end{figure}

\begin{figure}[t!]
    \centering
    \subfloat[Step 144: reached suboptimal distribution for the 1st budget ]{
        \includegraphics[width=0.45\textwidth]{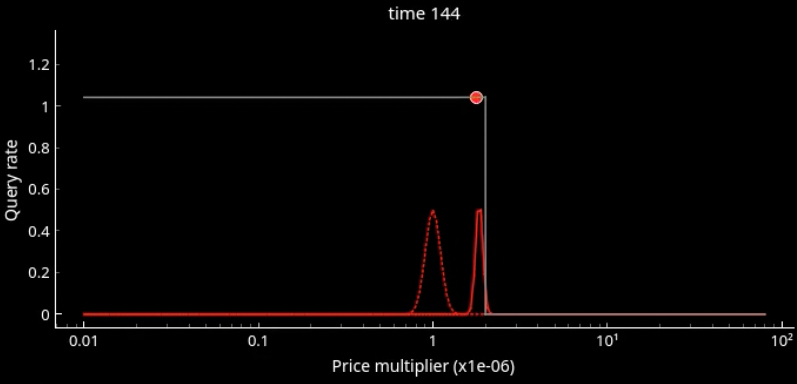}
        \label{fig:1ppo_dynamic_budget_144}
    }\\
    \subfloat[Step 201: consumer budget changed]{
        \includegraphics[width=0.45\textwidth]{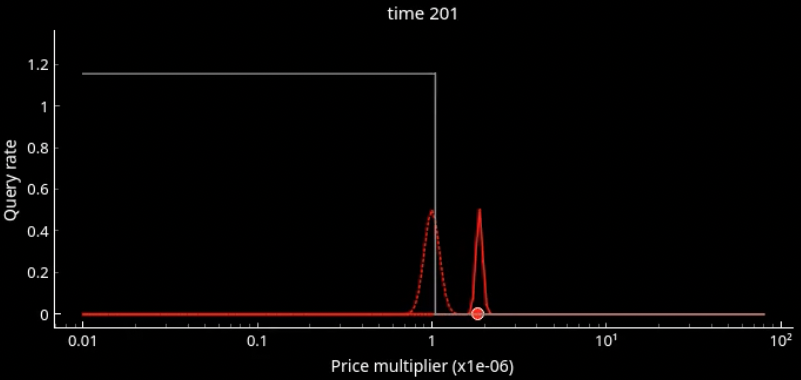}
        \label{fig:1ppo_dynamic_budget_201}
    }\\
    \subfloat[Step 251: spread of policy distribution]{
        \includegraphics[width=0.45\textwidth]{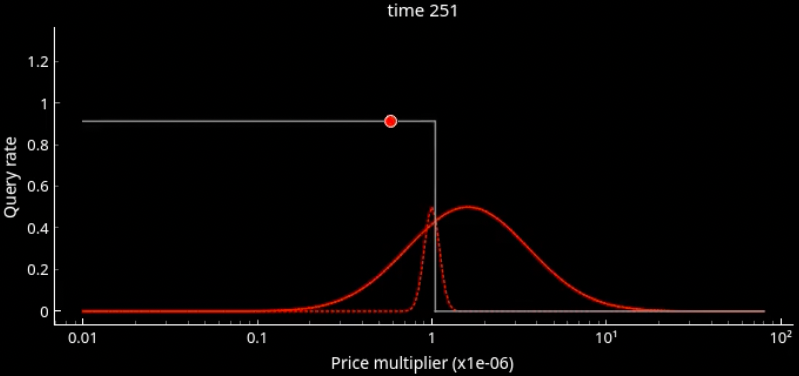}
        \label{fig:1ppo_dynamic_budget_251}
    }\\
    \subfloat[Step 398: reached suboptimal distribution for new budget]{
        \includegraphics[width=0.45\textwidth]{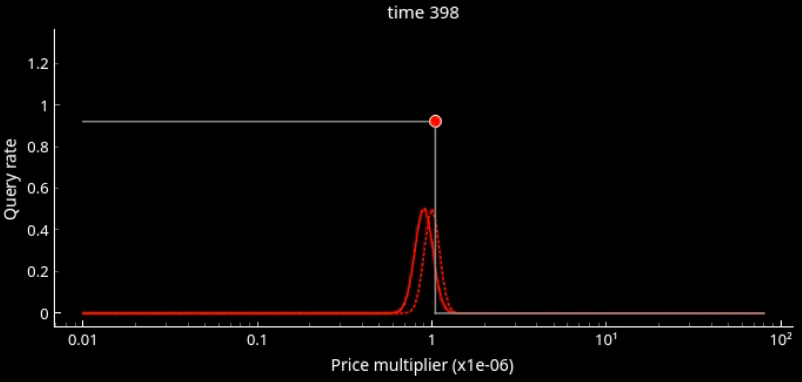}
        \label{fig:1ppo_dynamic_budget_398}
    }\\
    \caption{Experiments with dynamic Consumer budget and no competition.
    The budget changes rapidly in (b), hence the (once) suboptimal policy (b) first widens to sample from larger distribution (c) to finally converge to another suboptimal distribution (d).}
    \label{fig:1ppo_dynamic_budget}
\end{figure}

\begin{figure}[t!]
    \centering
    \subfloat[Step 198: reached suboptimal distribution]{
        \includegraphics[width=0.45\textwidth]{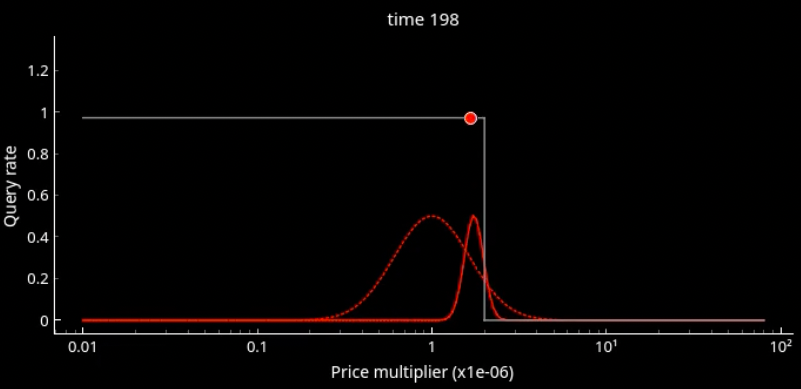}
        \label{fig:1ppo_no_demand_198} 
    }\\
    \subfloat[Step 200: demand for queries ceases]{
        \includegraphics[width=0.45\textwidth]{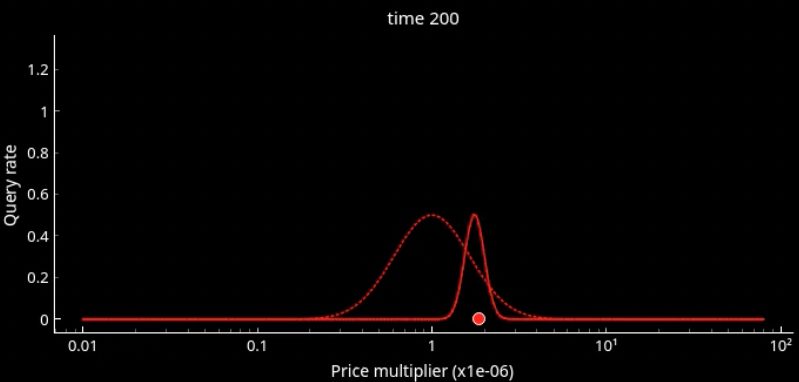}
        \label{fig:1ppo_no_demand_200}
        
    }\\
    \subfloat[Step 399: distribution is pulled towards init policy distribution]{
        \includegraphics[width=0.45\textwidth]{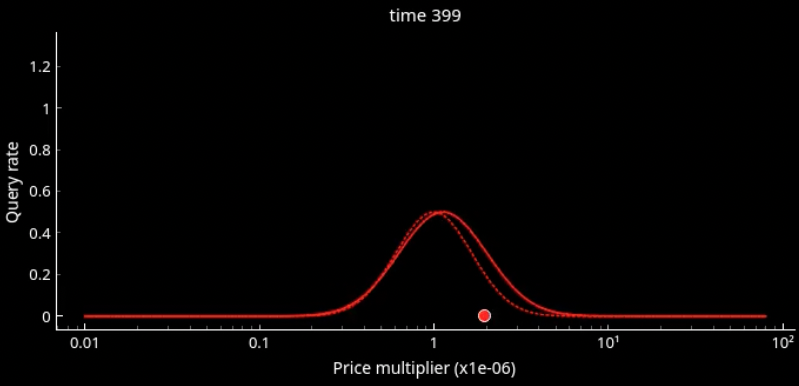}
        \label{fig:1ppo_no_demand_399} 
    }\\
    \caption{Experiments where query demand ceases (b) and bandit activates a mechanism pulling its distribution towards the initial one (c).}
    \label{fig:1ppo_no_demand}
\end{figure}

As a reminder, we assume that the cost to serve a query is 0.
Here, $\mathbbm{1}\{\textrm{condition}\}$ is an indicator function that returns 1 if the condition is true and 0 if it is false.
The agent, of course, tries to maximize its expected return over the entire game.

The problem with this is that $\textbf{b}_t$ is private information -- the agent has no way to know it.
Thus, it is impossible for the agent to actually optimize $r_t$ by direct optimization.
Instead, what the agent can observe, is $r_t$ given $p_t$.

This last observation is of key importance.
Our agent only has access to the reward.
Thus, a natural choice for the type of agent would be to use a bandit \cite{berry1985bandit, lattimore2020bandit} since bandits do not have an internal representation of the environment.
More specifically, for these, we use a Gaussian process bandit \cite{srinivas2009gaussian}.
The learning process for the bandit is as follows:
The agent samples a price from a Gaussian policy.
If the ISA does not reward the agent, either the price was too high or there are no queries to be served.
In both cases, the Gaussian widens so as to sample a greater variety of prices to try to find rewards.
If the agent receives a query below its mean price, then the policy's mean is decreased.
If the agent receives a query above its mean price, then the policy's mean is increased.
Finally, if the agent receives queries consistently, then the policy's variance is decreased.

\subsection{Consumer Budget Discovery}

In the first set of experiments, we check the bandit's performance under the simplest market conditions: a fixed  budget and no competition.
\autoref{fig:1ppo_fixed_budget} shows the agent's policy in two different time steps: in \autoref{fig:1ppo_fixed_budget_init} the agent learned that the budget is higher from its bids, hence it increased both the mean and variance to sample from a wider distribution and higher prices, whereas in \autoref{fig:1ppo_fixed_budget_subopt} the agent converged to the policy and with high probability samples from a narrow distribution just below the Consumer threshold.
Throughout this paper we will refer to this distribution as \textit{suboptimal}.

Next, we experimented with a dynamic Consumer budget, presented in \autoref{fig:1ppo_dynamic_budget}.
Initially, the agent successfully managed to discover the Consumer budget (\autoref{fig:1ppo_dynamic_budget_144}).
Thereafter, the Consumer budget decreased (\autoref{fig:1ppo_dynamic_budget_201}), and the agent stopped receiving queries.
Once the policy widened and its mean was reduced, the agent began to receive queries again (\autoref{fig:1ppo_dynamic_budget_251}).
Finally, the policy converged to the new suboptimal distribution (\autoref{fig:1ppo_dynamic_budget_398}).

In the third set of experiments, we simulated the case where Consumers are not sending any queries.
This situation actually happens in the real world occasionally and  requires special handling. We modify the bandit's policy update rule by incorporating a small ``pull'' toward the initial distribution when its experience replay buffer is empty.

\autoref{fig:1ppo_no_demand_200} presents one of the experiments when the initial demand is suddenly gone, starting at step 200.
Once the agent detects that situation it changes its update rule and starts slowly converging toward the initial distribution (\autoref{fig:1ppo_no_demand_399}) and will remain there until query demand reappears.

\begin{figure}[t!]
    \centering
    \subfloat[Step 241: bandits' distribution shifting towards budget]{
        \includegraphics[width=0.45\textwidth]{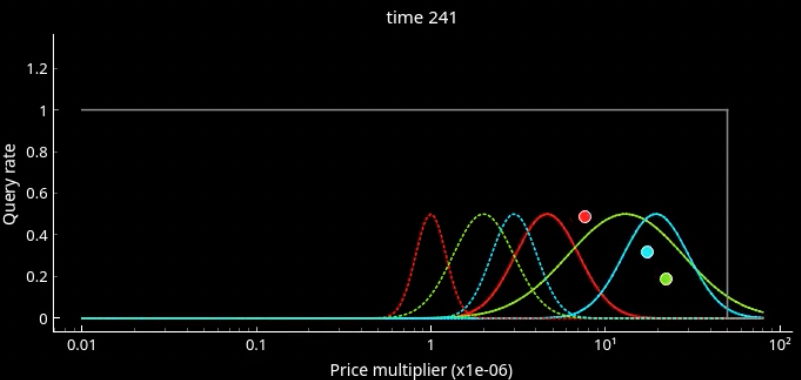}
        \label{fig:3ppo_policies_241} 
    }\\
    \subfloat[Step 999: bandits converged to suboptimal distributions]{
        \includegraphics[width=0.45\textwidth]{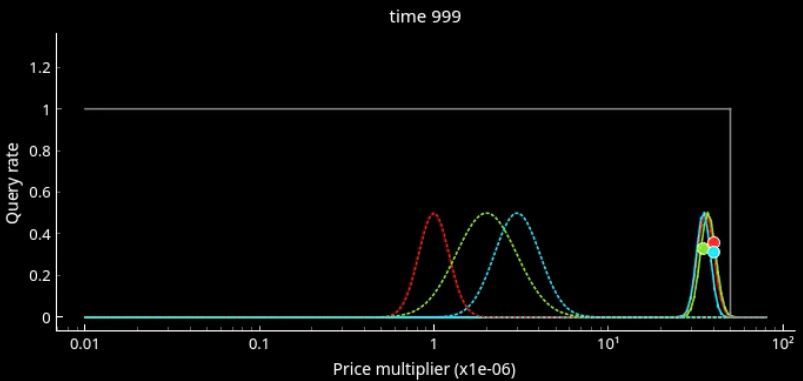}
        \label{fig:3ppo_policies_999} 
    }\\
    \subfloat[Total queries served by the bandits. The violet line indicates dropped queries (in this case: zero)]{
        \includegraphics[width=0.45\textwidth]{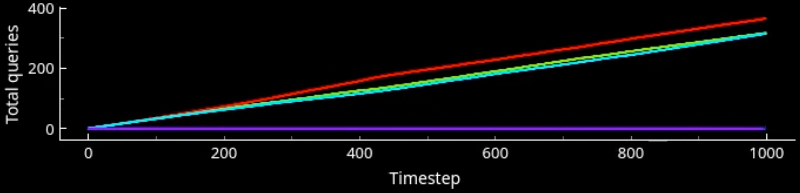}
        \label{fig:3ppo_policies_served} 
    }\\
    \subfloat[Total revenue of the three bandits]{
        \includegraphics[width=0.45\textwidth]{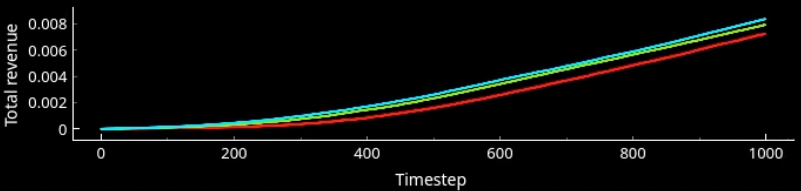}
        \label{fig:3ppo_policies_revenue} 
    }
    \caption{Experiments with multiple bandits competing for queries. The bandits distributions are moving (a) towards the suboptimal distributions (b).}
    \label{fig:3ppo_policies}
\end{figure}

\subsection{Multi-Agent Experiments}

We expected multiple Indexers to adopt our tool (a hypothesis that has since been borne out in reality), so we needed to ensure a satisfactory outcome when multiple agents competed.
Say we have $N$ agents competing for queries.
The $i$th agent selects some price $p^i_t \in \mathcal{P}^i$.
$\mathcal{P}^i$ represents the $i$th agent's minimum and maximum prices.
For a consumer query with budget $b^j_t$, the ISA will select the $i$th agent with some probability $P(p^i_t, b^j_t, \phi_t, p^{-i}_t)$, where $\phi_t$ are the unobservable features of all the bidders that the ISA takes into account when deciding which agent to select and $p^{-i}_t$ are the prices of all other agents, which are unobservable to agent $i$.
The $i$th agent's expected reward is given by

\begin{equation}
    \label{eq:marl-reward}
    R^i_t(p^i_t, p^{-i}_t, \textbf{b}_t, \phi_t) = \sum_{j} p^i_t P(p^i_t, b^j_t, \phi_t, p^{-i}_t)
\end{equation}

We model this as a Markov game.
Crucially, the probability that the ISA selects agent $i$, and thus the probability that agent $i$ is rewarded for a given query, depends on the joint-action $p_t$ -- a vector of prices in which the $i$th element is $p^i_t$.
Thus, we can re-write \autoref{eq:marl-reward} as

\begin{equation}
    \label{eq:marl-reward-simple}
    R^i_t(p^i, \textbf{b}_t, \phi_t) = \sum_{j} p^i_t P(p_t, b^j_t, \phi_t)
\end{equation}

The multi-agent setting also suffers from unobservable features in the environment.
To make matters more complex, those features are actively optimized by competitors, meaning that there is an added degree of non-stationarity.
Here, we wouldn't expect the bandit to perform well by itself.
We will discuss this topic more in the next section devoted to the impact of the Gateway.

\begin{figure}[t!]
    \centering
    \subfloat[Step 0: bandits starting from the same init distribution]{
        \includegraphics[width=0.45\textwidth]{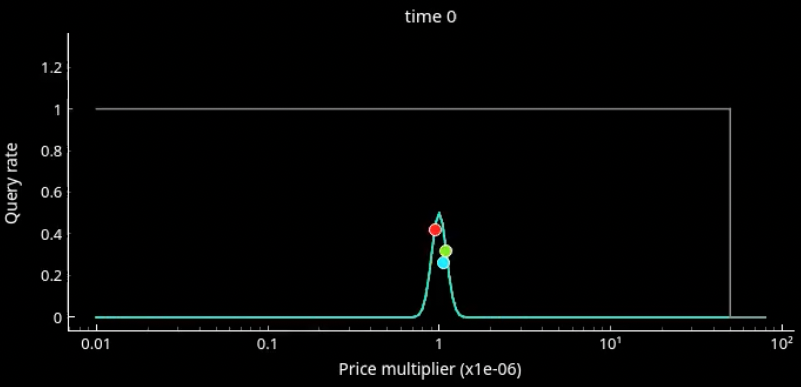}
        \label{fig:3bandits_policies_0}
    }\\
    \subfloat[Step 99: bandit's distributions shifts towards the budget]{
        \includegraphics[width=0.45\textwidth]{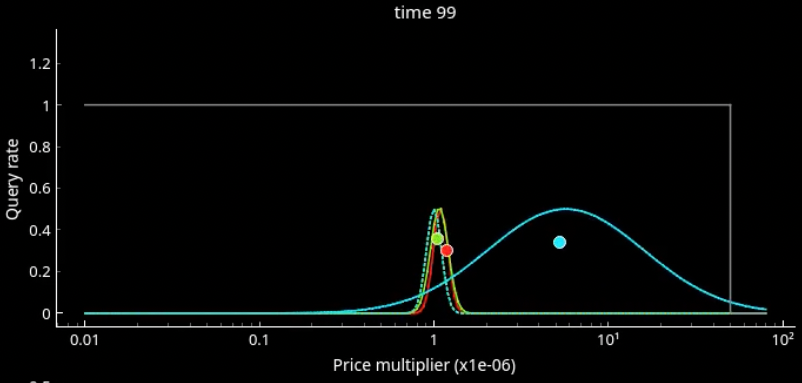}
        \label{fig:3bandits_policies_99}
    }\\
    \subfloat[Step 999: bandit's distribution continue shifting]{
        \includegraphics[width=0.45\textwidth]{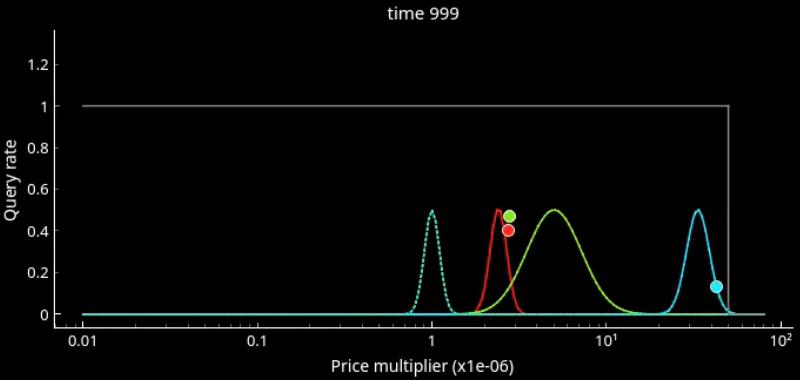}
        \label{fig:3bandits_policies_999}
    }\\
    \caption{Experiments with different bandits: red (vanilla Policy Gradients), green (PPO), cyan (our modified PPO).}
    \label{fig:3bandits_policies}
\end{figure}

In \autoref{fig:3ppo_policies}, we present an experiment with three Gaussian bandits competing for the same queries with a fixed Consumer budget.
All bandits utilized the same policy update rules, but started from different initial distributions.
As presented in (\autoref{fig:3ppo_policies_999}) all agents converged to the same suboptimal distribution.
Still, the total number of served queries (\autoref{fig:3ppo_policies_served}) and 
total revenue (\autoref{fig:3ppo_policies_revenue}) differ.
In particular, the ``red bandit'' that started with the lowest initial mean price 
was the most competitive and managed to serve more queries, but as it was serving for the lower price, hence ultimately it ended with the lowest revenue.
Similar effects were observed in experiments with more agents and different initial distributions.

Next, we conducted several experiments comparing the performance of bandits utilizing different policy update rules with a fixed Consumer budget.
We have compared bandits with three different update rules,
with \autoref{fig:3bandits_policies} showing how their policies evolved.
The red bandit used vanilla policy gradients~\cite{sutton1999policy,kakade2001natural} and the green used Proximal Policy Optimization (PPO)~\cite{schulman2017proximal}.
The cyan bandit employed a modified PPO update rule and the results clearly indicate that this algorithm is more sample-efficient and as a result the bandit discovers the Consumer budget faster.

It is worth noting that we have actually used the modified PPO rule in all the previous experiments whenever referring to a bandit, but decided not to explain it as the modification only impacts the bandit's sample-efficiency.
In short, the original PPO performs updates once its experience replay buffer is full, then clears the buffer and the process repeats.
As a result, the buffer always contains data associated with the current policy, therefore it is an on-policy algorithm.
Our modification changes the logic with respect to how the experience replay buffer operates. Specifically, for the modified PPO update rule, we do not clear the buffer when it is full, rather we truncate it to a maximum size whenever a new sample is collected.
As a result, the modified PPO buffer sometimes contains samples from both the current policy and any other of the policies used in the past, therefore it is an off-policy algorithm.

\begin{figure}[t!]
    \centering
    \subfloat[Step 999: bandit's policy at the end of experiment]{
        \includegraphics[width=0.45\textwidth]{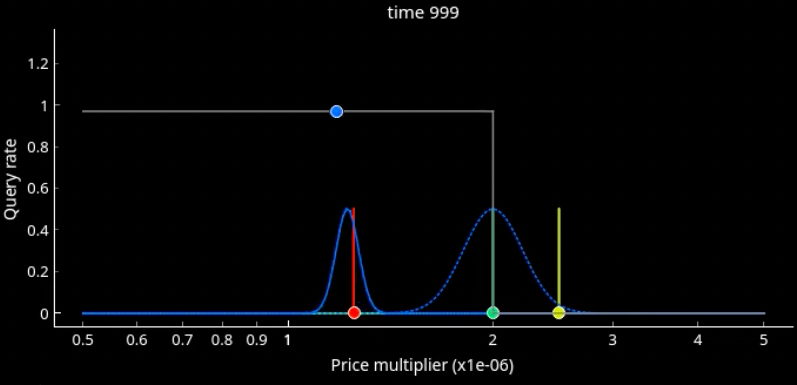}
        \label{fig:softmax_bandit_vs_deterministic_agents_999}
    }\\
    \subfloat[Queries served by the bandit and agents]{
        \includegraphics[width=0.45\textwidth]{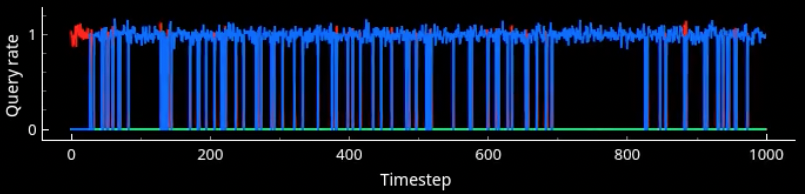}
        \label{fig:softmax_bandit_vs_deterministic_agents_served}
    }\\
    \subfloat[Total queries served by the bandits. Purple line indicates dropped queries (here: zero)]{
        \includegraphics[width=0.45\textwidth]{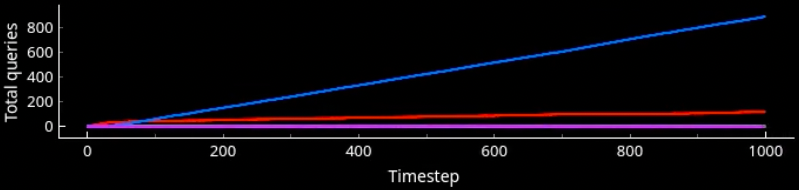}
        \label{fig:softmax_bandit_vs_deterministic_agents_total}
    }\\
    \caption{Experiments with environment implementing a naive query distribution algorithm and single Gaussian bandit (blue) competing with several deterministic agents with fixed prices (red, cyan, yellow).}
    \label{fig:softmax_bandit_vs_deterministic_agents}
\end{figure}

\begin{figure}[t]
    \centering
    \subfloat[Step 999: bandit's policy at the end of experiment]{
        \includegraphics[width=0.45\textwidth]{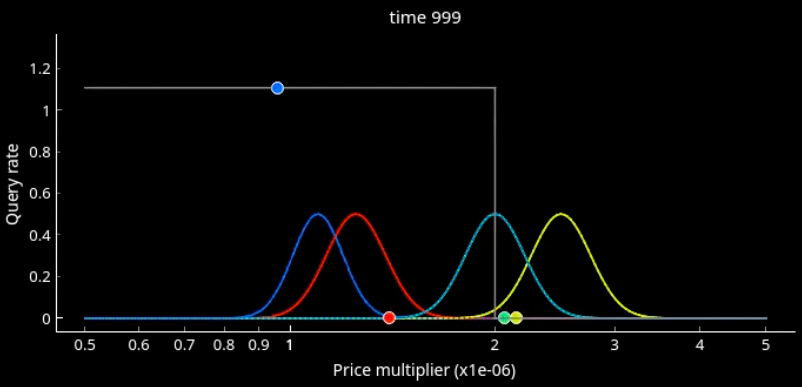}
        \label{fig:softmax_bandit_vs_stochastic_agents_999}
    }\\
    \subfloat[Queries served by the bandit and agents]{
        \includegraphics[width=0.45\textwidth]{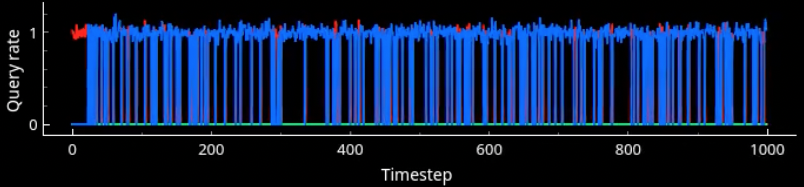}
        \label{fig:softmax_bandit_vs_stochastic_agents_served}
    }\\
    \subfloat[Total queries served by the bandit and agents. Purple line indicates dropped queries]{
        \includegraphics[width=0.45\textwidth]{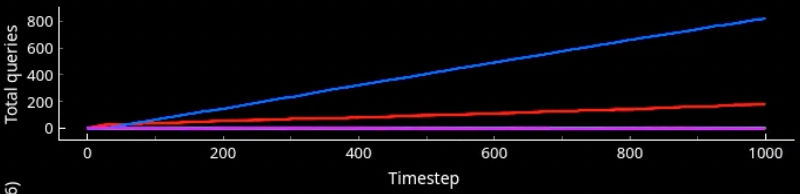}
        \label{fig:softmax_bandit_vs_stochastic_agents_total}
    }\\
    \caption{Experiments with environment implementing a naive Query distribution algorithm and single Gaussian bandit (blue) competing with several stochastic agents with fixed price distributions (red, cyan, yellow).}
    \label{fig:softmax_bandit_vs_stochastic_agents}
\end{figure}

\subsection{On the Impact of Gateways}
\label{sec:gateway_impact}



In the experiments presented above, the agents converged to the common suboptimal price.
One can notice that there is nothing special implemented in the agent policies that supports this.
Even more, the fact that their rewards are purely associated with the query volume should result in behavior that is competitive rather than collaborative.

It appears that the observed outcome actually results from the policy implemented inside the Gateway’s ISA that has to continuously probe and evaluate the Indexers serving the queries.
To investigate this further, we have performed several multi-agent experiments with a modified environment where we replaced the ISA with a naive Query distribution algorithm that splits the total volume between agents inverse proportionally to their bids.

In \autoref{fig:softmax_bandit_vs_deterministic_agents} we present results where a single Gaussian bandit competed against three deterministic agents with fixed policies.
After initially overshooting the price, the Gaussian bandit subsequently decreases its price successfully discovers the price bids of the fixed agents (\autoref{fig:softmax_bandit_vs_deterministic_agents_999}).
As a result, the Gaussian bandit dominates the competition and finds a suboptimal policy with a mean just below the price that enables it to capture the majority of query volume, see (Figures \ref{fig:softmax_bandit_vs_deterministic_agents_served} and \ref{fig:softmax_bandit_vs_deterministic_agents_total}).
Note that drops in (\autoref{fig:softmax_bandit_vs_deterministic_agents_served}) result from the Gaussian bandit exploring to check if the price can be increased--and in doing so losing the bidding to the red agent.

\begin{figure}[t!]
    \centering
    \subfloat[Bandit policies at step 99]{
        \includegraphics[width=0.45\textwidth]{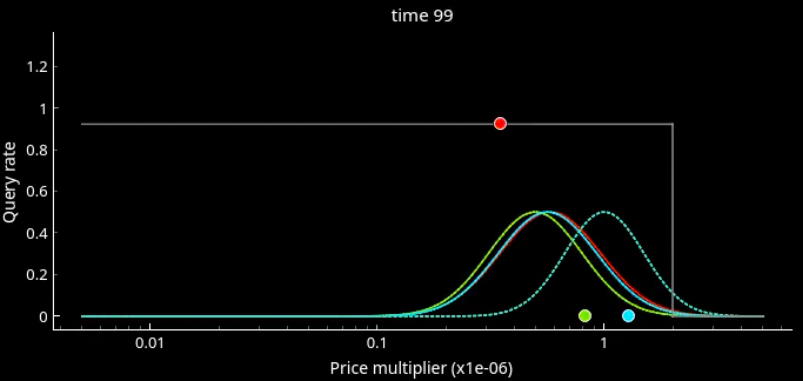}
        \label{fig:softmax_3bandits_99}
    }\\
    \subfloat[Bandit policies at step 499]{
        \includegraphics[width=0.45\textwidth]{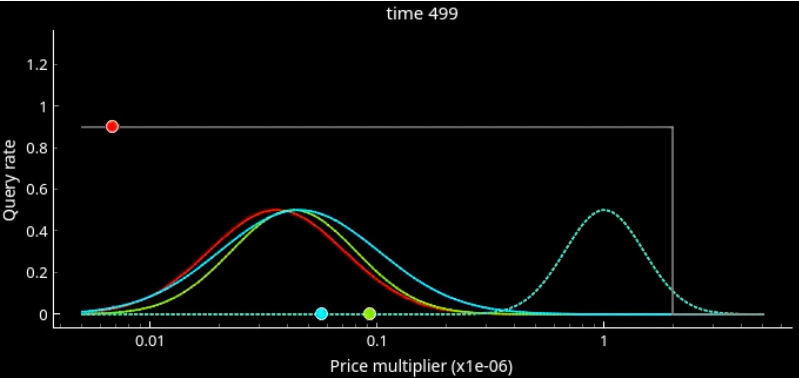}
        \label{fig:softmax_3bandits_499}
    }\\
    \subfloat[Total queries served by the bandits. The violet line indicates dropped queries (in this case: zero)]{
        \includegraphics[width=0.45\textwidth]{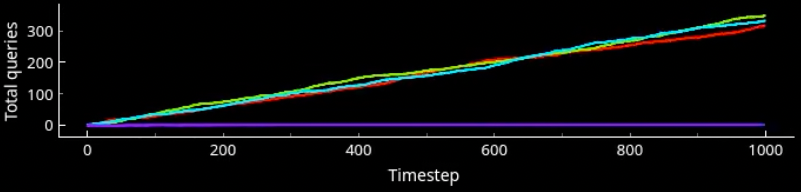}
        \label{fig:softmax_3bandits_total_queries}
    }\\
    \subfloat[Bandit revenue rates]{
        \includegraphics[width=0.45\textwidth]{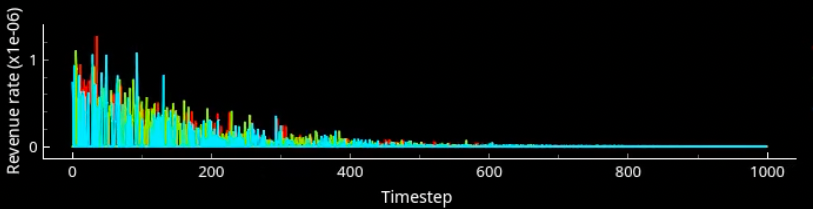}
        \label{fig:softmax_3bandits_revenue_rate}
    }\\
    \subfloat[Total bandit revenues]{
        \includegraphics[width=0.45\textwidth]{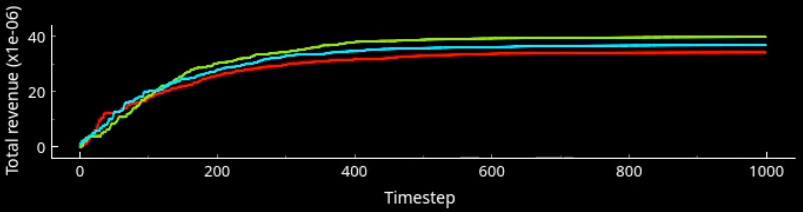}
        \label{fig:softmax_3bandits_total_revenue} 
    }\\
    \caption{Experiments with environment implementing a naive Query distribution algorithm and three Gaussian bandits competing with each other.}
    \label{fig:softmax_3bandits}
\end{figure}

A very similar situation happens when the bandit competes with stochastic agents with fixed policy distributions (\autoref{fig:softmax_bandit_vs_stochastic_agents}.
The main difference is that the final bandit's policy distribution is wider, which is a natural consequence of competing with the stochastic policy of another agent.

Finally, we ran several experiments with several bandits competing with each other (\autoref{fig:softmax_3bandits}).
As expected, in a non-regulated market and with bandits driven purely by the served volume of queries, the bandits started to fight for market domination by slowly decreasing their prices.
This might be perceived as a classical tit for tat strategy~\cite{axelrod1981evolution} as one bandit drops the price a bit as a reaction to the other bandit dropping the price earlier.

This results in a race-to-the-bottom outcome (compare Figures \ref{fig:softmax_3bandits_99} and \ref{fig:softmax_3bandits_499}) with bandits' revenue asymptotically dropping to zero (\autoref{fig:softmax_3bandits_revenue_rate}).
This is a negative outcome, as, in the long run, Indexers using such policies would abandon subgraphs that do not bring revenue, and a single Indexer could then enter the market to monopolize queries.
This outcome is exactly what The Graph is trying to avoid.




\section{Field Experiments}

We have deployed into production a solution based on the Gaussian bandit with our modified PPO policy update rule. The deployed bandit is controlling the price bids for our Indexer operating on the Ethereum blockchain.
In the span of more than 12 hours, the Indexer was actually serving queries for several subgraphs, with our bandit controlling its query price.
The Graph's Indexer software was executing additional logic for logging served queries and aggregating them into query volume that every three minutes were used as a reward signal for the bandit to update its policy.
Additionally, we collected statistics on the bandit's mean price and its standard deviation, as well as estimated our Indexer's revenue.

\begin{figure}[t!]
    \centering
    \subfloat[Bandit's policy mean]{
        \includegraphics[width=0.5\textwidth]{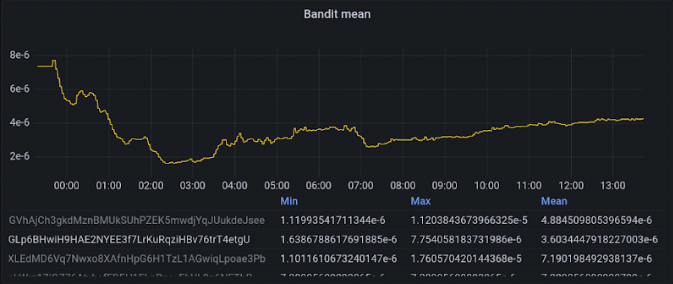}
        \label{fig:bandit_in_production_mean}
    }\\
    \subfloat[Bandit's policy standard deviation]{
        \includegraphics[width=0.5\textwidth]{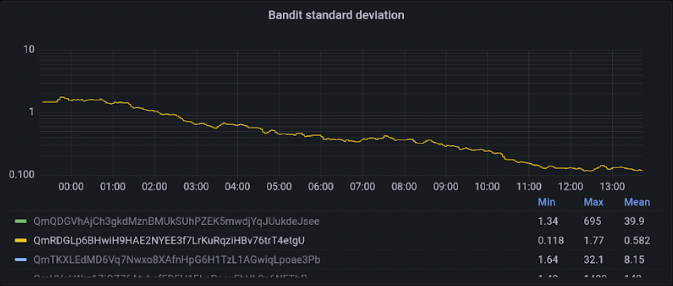}
        \label{fig:bandit_in_production_stddev}
    }\\
    \subfloat[Bandit's revenue rate]{
        \includegraphics[width=0.5\textwidth]{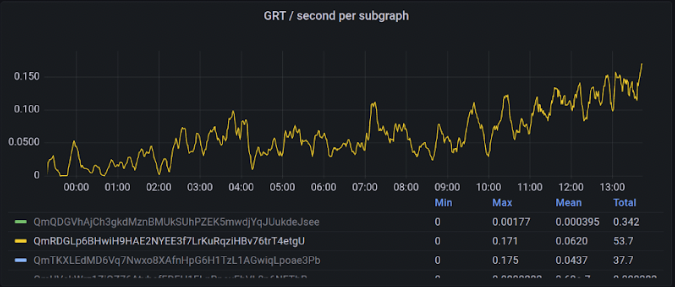}
        \label{fig:bandit_in_production_revenue}
    }\\
    \caption{Field experiments with bandit deployed in production on one of The Graph's Indexers}
    \label{fig:bandit_in_production}
\end{figure}

Note that the traffic and customers' budgets were changing continuously.
Still, after the initial phase where the bandit overshot the mean price, it started to slowly increase its mean (\autoref{fig:bandit_in_production_mean}) and decrease its standard deviation (\autoref{fig:bandit_in_production_stddev}).
Moreover, there is a clear tendency for the revenue that was increasing during consecutive periods (\autoref{fig:bandit_in_production_revenue}).

The above indicates that there is a positive impact of the bandit. However, currently, we cannot show any quantitative results.
First, there is no baseline that we could compare to--an Indexer in production is responding to demand originating from multiple Consumers submitting queries with constantly fluctuating demand.
Moreover, The Graph currently lacks proper tooling for measuring the impact of our solution on the scale of the entire distributed network of Indexers.



\section{Conclusions and Future Work}

In this paper, we have developed a bandit-based solution for dynamic pricing in a decentralized marketplace, and we have deployed it to production on a blockchain protocol.
As certain subgraphs have very low query volume, we endowed our bandit with an additional mechanism that pulls its policy toward its initial policy distribution to prevent its variance from approaching infinity when it does not serve queries for a long time.
We also improved the sample efficiency of our Gaussian bandits by modifying the experience replay buffer used for its PPO update rule.

While this tool is now in production for multiple competing Indexers, we need to further our algorithmic development and our understanding of the market impact of our agents.
We plan to build tooling that Indexers can optionally run to collect and send us data on their revenue, enabling us to understand the real-world impact beyond just our own Indexer.
This data currently only exists in aggregate figures.

Moreover, neither the simulator nor the tool yet cover the full scope of constraints that we would like them to.
For example, Indexers have hardware constraints and cannot serve an unlimited number of queries at the same time.
Hence after serving some number of queries, our algorithm may want to intentionally set a higher price-bid to reduce the number of queries the Indexer receives, while still ensuring the Indexer is earning revenue and maintaining an acceptable level of Quality of Service.

We also wish to further our investigation on both sample efficiency and the decentralized multi-agent optimization problems.
In particular, the ISA is an area of active research within The Graph.
We hope this work and the associated open-source tools for simulation will assist The Graph ecosystem in developing the future versions of ISA.

\section{Acknowledgments}
The authors acknowledge the support of The Graph Foundation.
This material is based upon work made under The Graph Foundation grant for Core Developers.
The authors would like to thank their collaborators, in particular, Zac Burns and Theo Butler from Edge \& Node, for constructive feedback and discussions that led to the development of the solutions presented in this paper.

\bibliography{aaai22.bib}

\begin{thebibliography}{29}
\providecommand{\natexlab}[1]{#1}

\bibitem[{Axelrod and Hamilton(1981)}]{axelrod1981evolution}
Axelrod, R.; and Hamilton, W.~D. 1981.
\newblock The evolution of cooperation.
\newblock \emph{science}, 211(4489): 1390--1396.

\bibitem[{Berry and Fristedt(1985)}]{berry1985bandit}
Berry, D.~A.; and Fristedt, B. 1985.
\newblock Bandit problems: sequential allocation of experiments (Monographs on
  statistics and applied probability).
\newblock \emph{London: Chapman and Hall}, 5(71-87): 7--7.

\bibitem[{Buterin(2014)}]{buterin2014next}
Buterin, V. 2014.
\newblock A next-generation smart contract and decentralized application
  platform.
\newblock \emph{White paper}, 3(37): 2--1.

\bibitem[{Cheung, Simchi-Levi, and Wang(2017)}]{cheung2017dynamic}
Cheung, W.~C.; Simchi-Levi, D.; and Wang, H. 2017.
\newblock Dynamic pricing and demand learning with limited price
  experimentation.
\newblock \emph{Operations Research}, 65(6): 1722--1731.

\bibitem[{{Facebook, Inc.}(2015)}]{graphql2015spec}
{Facebook, Inc.} 2015.
\newblock RFC Specification for GraphQL.

\bibitem[{Foerster et~al.(2017)Foerster, Nardelli, Farquhar, Afouras, Torr,
  Kohli, and Whiteson}]{foerster2017stabilising}
Foerster, J.; Nardelli, N.; Farquhar, G.; Afouras, T.; Torr, P.~H.; Kohli, P.;
  and Whiteson, S. 2017.
\newblock Stabilising experience replay for deep multi-agent reinforcement
  learning.
\newblock In \emph{International conference on machine learning}, 1146--1155.
  PMLR.

\bibitem[{Hartig and P{\'e}rez(2018)}]{hartig2018semantics}
Hartig, O.; and P{\'e}rez, J. 2018.
\newblock Semantics and complexity of GraphQL.
\newblock In \emph{Proceedings of the 2018 World Wide Web Conference},
  1155--1164.

\bibitem[{Jain et~al.(1984)Jain, Chiu, Hawe et~al.}]{jain1984quantitative}
Jain, R.~K.; Chiu, D.-M.~W.; Hawe, W.~R.; et~al. 1984.
\newblock A quantitative measure of fairness and discrimination.
\newblock \emph{Eastern Research Laboratory, Digital Equipment Corporation,
  Hudson, MA}, 21.

\bibitem[{Kakade(2001)}]{kakade2001natural}
Kakade, S.~M. 2001.
\newblock A natural policy gradient.
\newblock \emph{Advances in neural information processing systems}, 14.

\bibitem[{K{\"o}n{\"o}nen(2006)}]{kononen2006dynamic}
K{\"o}n{\"o}nen, V. 2006.
\newblock Dynamic pricing based on asymmetric multiagent reinforcement
  learning.
\newblock \emph{International journal of intelligent systems}, 21(1): 73--98.

\bibitem[{Lattimore and Szepesv{\'a}ri(2020)}]{lattimore2020bandit}
Lattimore, T.; and Szepesv{\'a}ri, C. 2020.
\newblock \emph{Bandit algorithms}.
\newblock Cambridge University Press.

\bibitem[{Liu et~al.(2019)Liu, Zhang, Wang, Deng, and Wu}]{liu2019dynamic}
Liu, J.; Zhang, Y.; Wang, X.; Deng, Y.; and Wu, X. 2019.
\newblock Dynamic Pricing on E-commerce Platform with Deep Reinforcement
  Learning: A Field Experiment.
\newblock \emph{arXiv preprint arXiv:}.

\bibitem[{Maestre et~al.(2018)Maestre, Duque, Rubio, and
  Ar{\'e}valo}]{maestre2018reinforcement}
Maestre, R.; Duque, J.; Rubio, A.; and Ar{\'e}valo, J. 2018.
\newblock Reinforcement learning for fair dynamic pricing.
\newblock In \emph{Proceedings of SAI Intelligent Systems Conference},
  120--135. Springer.

\bibitem[{Matignon, Laurent, and Le~Fort-Piat(2012)}]{matignon2012independent}
Matignon, L.; Laurent, G.~J.; and Le~Fort-Piat, N. 2012.
\newblock Independent reinforcement learners in cooperative markov games: a
  survey regarding coordination problems.
\newblock \emph{The Knowledge Engineering Review}, 27(1): 1--31.

\bibitem[{Mavroudeas et~al.(2021)Mavroudeas, Baudart, Cha, Hirzel, Laredo,
  Magdon-Ismail, Mandel, and Wittern}]{mavroudeas2021learning}
Mavroudeas, G.; Baudart, G.; Cha, A.; Hirzel, M.; Laredo, J.~A.; Magdon-Ismail,
  M.; Mandel, L.; and Wittern, E. 2021.
\newblock Learning GraphQL query cost.
\newblock In \emph{2021 36th IEEE/ACM International Conference on Automated
  Software Engineering (ASE)}, 1146--1150. IEEE.

\bibitem[{Nakhe(2017)}]{nakhe2017dynamic}
Nakhe, P. 2017.
\newblock Dynamic pricing in competitive markets.
\newblock In \emph{International Conference on Web and Internet Economics},
  354--367. Springer.

\bibitem[{Saberi et~al.(2019)Saberi, Kouhizadeh, Sarkis, and
  Shen}]{saberi2019blockchain}
Saberi, S.; Kouhizadeh, M.; Sarkis, J.; and Shen, L. 2019.
\newblock Blockchain technology and its relationships to sustainable supply
  chain management.
\newblock \emph{International Journal of Production Research}, 57(7):
  2117--2135.

\bibitem[{Schulman et~al.(2017)Schulman, Wolski, Dhariwal, Radford, and
  Klimov}]{schulman2017proximal}
Schulman, J.; Wolski, F.; Dhariwal, P.; Radford, A.; and Klimov, O. 2017.
\newblock Proximal policy optimization algorithms.
\newblock \emph{arXiv preprint arXiv:1707.06347}.

\bibitem[{Srinivas et~al.(2009)Srinivas, Krause, Kakade, and
  Seeger}]{srinivas2009gaussian}
Srinivas, N.; Krause, A.; Kakade, S.~M.; and Seeger, M. 2009.
\newblock Gaussian process optimization in the bandit setting: No regret and
  experimental design.
\newblock \emph{arXiv preprint arXiv:0912.3995}.

\bibitem[{Sutton and Barto(2018)}]{sutton2018reinforcement}
Sutton, R.~S.; and Barto, A.~G. 2018.
\newblock \emph{Reinforcement learning: An introduction}.
\newblock MIT press.

\bibitem[{Sutton et~al.(1999)Sutton, McAllester, Singh, and
  Mansour}]{sutton1999policy}
Sutton, R.~S.; McAllester, D.; Singh, S.; and Mansour, Y. 1999.
\newblock Policy gradient methods for reinforcement learning with function
  approximation.
\newblock \emph{Advances in neural information processing systems}, 12.

\bibitem[{Tan(1993)}]{tan1993multi}
Tan, M. 1993.
\newblock Multi-agent reinforcement learning: Independent vs. cooperative
  agents.
\newblock In \emph{Proceedings of the tenth international conference on machine
  learning}, 330--337.

\bibitem[{Tapscott and Tapscott(2017)}]{tapscott2017blockchain}
Tapscott, A.; and Tapscott, D. 2017.
\newblock How blockchain is changing finance.
\newblock \emph{Harvard Business Review}, 1(9): 2--5.

\bibitem[{Tesauro and Kephart(2002)}]{tesauro2002pricing}
Tesauro, G.; and Kephart, J.~O. 2002.
\newblock Pricing in agent economies using multi-agent Q-learning.
\newblock \emph{Autonomous agents and multi-agent systems}, 5(3): 289--304.

\bibitem[{{The Graph Foundation}(2022)}]{thegraph2022}
{The Graph Foundation}. 2022.
\newblock An Introduction to Web3 and The Graph for New Users.
\newblock \url{https://thegraph.com/blog/introduction-to-the-graph/}.

\bibitem[{Third and Domingue(2017)}]{third2017linked}
Third, A.; and Domingue, J. 2017.
\newblock Linked data indexing of distributed ledgers.
\newblock In \emph{Proceedings of the 26th International Conference on World
  Wide Web Companion}, 1431--1436.

\bibitem[{Wood(2014)}]{wood2014ethereum}
Wood, G. 2014.
\newblock Ethereum: A secure decentralized transaction ledger.
\newblock \url{https://gavwood.com/Paper.pdf}.

\bibitem[{Zhang, Yang, and Ba{\c{s}}ar(2021)}]{zhang2021multi}
Zhang, K.; Yang, Z.; and Ba{\c{s}}ar, T. 2021.
\newblock Multi-agent reinforcement learning: A selective overview of theories
  and algorithms.
\newblock \emph{Handbook of Reinforcement Learning and Control}, 321--384.

\bibitem[{Zheng et~al.(2018)Zheng, Xie, Dai, Chen, and
  Wang}]{zheng2018blockchain}
Zheng, Z.; Xie, S.; Dai, H.-N.; Chen, X.; and Wang, H. 2018.
\newblock Blockchain challenges and opportunities: A survey.
\newblock \emph{International journal of web and grid services}, 14(4):
  352--375.

\end{thebibliography}
\end{document}